\documentclass[12pt]{article}
\usepackage{latexsym}
\usepackage{amssymb}
\usepackage{graphicx}
\usepackage{cite}

\newtheorem{Theorem}{Theorem}[part]

\newtheorem{Proposition}{Proposition}[part]
\newtheorem{Assumption}{Assumption}[part]
\newtheorem{Lemma}{Lemma}[part]
\newtheorem{Corollary}{Corollary}[part]
\newtheorem{Remark}{Remark}[part]
\newtheorem{Example}{Example}[part]
\newtheorem{Algorithm}{Algorithm}

\def \ep{\hbox{ }\hfill$\Box$}

\def\reff#1{{\rm(\ref{#1})}}

\begin{document}
\title{The geometric measure of entanglement of pure states with nonnegative amplitudes and the spectral theory of
nonnegative tensors}

\author{
Shenglong Hu \thanks{Email: Tim.Hu@connect.polyu.hk. Department of
Applied Mathematics, The Hong Kong Polytechnic University, Hung Hom,
Kowloon, Hong Kong.},\hspace{4mm}\hspace{4mm}Liqun Qi \thanks{Email:
maqilq@polyu.edu.hk. Department of Applied Mathematics, The Hong
Kong Polytechnic University, Hung Hom, Kowloon, Hong Kong. This
author's work was supported by the Hong Kong Research Grant
Council.},\hspace{4mm}\hspace{4mm} Guofeng Zhang \thanks{Email:
magzhang@inet.polyu.edu.hk. Department of Applied Mathematics, The
Hong Kong Polytechnic University, Hung Hom, Kowloon, Hong Kong. This
author's work was supported by the Hong Kong Research Grant
Council.}}

\date{\today} \maketitle

\begin{abstract}
\noindent The geometric measure of entanglement for a symmetric pure
state with nonnegative amplitudes has attracted much attention.  On the
other hand, the spectral theory of nonnegative tensors
(hypermatrices) has been developed rapidly. In this paper, we show
how the spectral theory of nonnegative tensors can be applied to the
study of the geometric measure of entanglement for a pure state with
nonnegative amplitudes. Especially, an elimination method for
computing the geometric measure of entanglement for symmetric pure
multipartite qubit or qutrit states with nonnegative amplitudes is
given. For symmetric pure multipartite qudit states with nonnegative
amplitudes, a numerical algorithm with randomization is presented
and proven to be convergent. We show that for the geometric measure of
entanglement for pure states with nonnegative
amplitudes, the nonsymmetric ones can be converted to the symmetric ones.

 \vspace{3mm}

\noindent {\bf Key words:}\hspace{2mm} quantum entanglement, geometric measure, Z-eigenvalue, nonnegative tensor
\vspace{3mm}
\end{abstract}


\section{Introduction}

\setcounter{Assumption}{0}
\setcounter{Theorem}{0} \setcounter{Proposition}{0}
\setcounter{Corollary}{0} \setcounter{Lemma}{0}
\setcounter{Definition}{0} \setcounter{Remark}{0}
\setcounter{Algorithm}{0}  \setcounter{Example}{0}

\hspace{4mm} The quantum entanglement problem is regarded as a central problem in quantum information \cite{nc00}, and the geometric measure is one of the most important measures of quantum entanglement
\cite{bh01,s95,wg03,hkwg09,hmmov09,cxz10,odv08}. It was first proposed by Shimony \cite{s95} and generalized to multipartite systems by Wei and Goldbart \cite{wg03}, and has become one of the widely used entanglement measures for multiparticle cases \cite{hkwg09,hmmov09,cxz10,q122}. Among them, the study on pure states with nonnegative amplitudes attracted much attention recently. Wei and Goldbart conjectured in
\cite[Section II.A]{wg03} that the nearest separable state for a symmetric state can be chosen to be symmetric. Hayashi, Markham, Murao, Owari and Virmani \cite{hmmov09} proved the conjecture for the special case of symmetric states with nonnegative amplitudes and that the nearest separable state can be chosen with nonnegative amplitudes in this situation, and H\"{u}bener, Kleinmannn, Wei, Gonz\'{a}lez-Guill\'{e}n and G\"{u}hne \cite{hkwg09} proved the conjecture completely. The computation of the symmetric pure states with nonnegative amplitudes was carried out by Wei and Goldbart \cite{wg03} for some ground states, and systematically for symmetric pure multipartite qubit states by Chen, Xu and Zhu \cite{cxz10}.
For general evaluations of the geometric entanglement for symmetric pure states, please see Or\'{u}s, Dusuel and Vidal \cite{odv08}, Chen, Xu and Zhu \cite{cxz10}, and references therein.

The central problem of
the computation of the geometric measure is to find the largest
entanglement eigenvalue \cite{hs10,q122,wg03}. Mathematically, the quantum eigenvalue
problem is a generalization of the singular value problem of a
complex matrix \cite{q122,hs10}. There have been several
generalizations of singular values / eigenvalues of matrices to
tensors (hypermatrices) recently \cite{q05,l05,cpz09}. These form the spectral
theory of tensors, please see \cite{q123} and references therein for the state of the art. In this paper, we investigate the geometric measure of pure states with nonnegative amplitudes through the spectral theory of nonnegative tensors.

More specially, we establish a connection between the concept of
Z-eigenvalues of tensors and the quantum eigenvalue problem. We show
that the geometric measure of a symmetric pure state with
nonnegative amplitudes is equal to the Z-spectral radius of the
underlying nonnegative tensor (Theorem \ref{thm-1}). Based on this
connection, a method in views of variable elimination
\cite{qww09,qyw10} for computing the geometric measure of
entanglement for symmetric pure multipartite qubit or qutrit states
with nonnegative amplitudes is given. For the qubit case, it is an
alternative to the method in \cite[Section II.A]{cxz10}. For the
qutrit case, it is new and gives an analytical derivation of the
geometric measure of entanglement for such states. Here a qutrit means a
system in a three dimensional Hilbert space, and a qudit used in
the sequel is for a higher dimeniaonl case, as in
\cite{cm99,rmndmc00,hj09}.

For symmetric pure multipartite qudit states with nonnegative
amplitudes cases, a numerical algorithm with randomization
(Algorithm \ref{algo}) is presented and proven to be convergent. The
method is based on the shifted higher order power method (Algorithm
\ref{algo1}) analyzed in \cite{km11}. We show that if the initial
points are randomly chosen from the intersection of the
positive orthant and the unit sphere, then with a positive
probability the algorithm finds the geometric measure of such states (Theorem \ref{thm-1-a}). The
probability of the convergence of the algorithm is determined by the
distance between the Z-spectral radius and the second largest
nonnegative Z-eigenvalue of the underlying tensor.

For nonsymmetric pure states with nonnegative amplitudes, similar results are established as well. Moreover, we show that for the geometric measure of
entanglement for pure states with nonnegative
amplitudes, the nonsymmetric ones can be converted to the symmetric ones. Consequently, the numerical methods established above are applicable.

The rest of this paper is organized as follows. The definitions and
some basic facts of the geometric measure and the Z-eigenvalues of
tensors are presented as preliminaries in the next section. In
Section 3, a connection between the geometric measure of a symmetric
pure state with nonnegative amplitudes and the theory of the
Z-spectral radius of a nonnegative tensor is established. The
computational issues are discussed. The details of the numerical
algorithm for multipartite qudit states are put in Appendix. In
Section 4, a connection between the geometric measure of a
nonsymmetric pure state with nonnegative amplitudes and the spectral
theory of nonnegative multilinear forms is established. The paper is
concluded with some final remarks in Section 5.

\section{Preliminaries}
\setcounter{Assumption}{0}
\setcounter{Theorem}{0} \setcounter{Proposition}{0}
\setcounter{Corollary}{0} \setcounter{Lemma}{0}
\setcounter{Definition}{0} \setcounter{Remark}{0}
\setcounter{Algorithm}{0}  \setcounter{Example}{0}

\hspace{4mm} In this section, some preliminaries of the geometric
measure of quantum entanglement and the Z-eigenvalues of tensors
(hypermatrices) are presented.

\subsection{Geometric measure}
An $m$-partite ($m\geq 3$ in this paper) pure state $|\Psi\rangle$ of a composite quantum system can be regarded as a normalized element in a
Hilbert tensor product space ${\mathcal{H}}=\bigotimes_{k=1}^m{\mathcal{H}}_k$, where the dimension of ${\mathcal{H}}_k$
is $d_k$ for $k=1,\ldots,m$. A separable $m$-partite state $|\Phi\rangle\in{\mathcal{H}}$ can be described by $|\Phi\rangle=\bigotimes_{k=1}^m|\phi^{(k)}\rangle$ with $|\phi^{(k)}\rangle\in{\mathcal{H}}_k$ and $\||\phi^{(k)}\rangle\|=1$ for $k=1,\ldots,m$. A state is called entangled if it is not separable. In this paper, only pure states are considered.

For a given $m$-partite pure state $|\Psi\rangle\in {\mathcal{H}}$, one
considers its nearest separable state
$|\Phi\rangle=\bigotimes_{k=1}^m|\phi^{(k)}\rangle$ in terms
of the maximal overlap:
\begin{eqnarray}\label{me}
G(\Psi)=\max_{|\Phi\rangle=\bigotimes_{k=1}^m|\phi^{(k)}\rangle}|\langle\Psi|\Phi\rangle|.
\end{eqnarray}
The geometric measure is then defined as \cite{wg03}
\begin{eqnarray*}
E_G(|\Psi\rangle)=1-G(\Psi)^2.
\end{eqnarray*}

It is shown that the maximal overlap in \reff{me} is equal to the largest entanglement eigenvalue $\lambda$ \cite{wg03,q122}:
\begin{eqnarray}\label{qe}
\left\{\begin{array}{rcl}\langle \Psi|\left(\bigotimes_{j\neq k}|\phi^{(j)}\rangle\right)&=&\lambda\langle\phi^{(k)}|,\\\left(\bigotimes_{j\neq k}\langle\phi^{(j)}|\right)\Psi\rangle&=&\lambda|\phi^{(k)}\rangle,\\\||\phi^{(k)}\rangle\|&=&1,k=1,\ldots,m.\end{array}\right.
\end{eqnarray}

A state
$|\Psi\rangle\in{\mathcal{H}}=\bigotimes_{k=1}^m{\mathcal{H}}_k$ is
called {\em nonnegative} if there exist orthonormal bases
$\{|e_i^{(k)}\rangle\}_{i=1}^{d_k}$ for ${\mathcal{H}}_k$ such that
$a_{i_1\ldots i_m}:=\langle\Psi|\left(|e_{i_1}^{(1)}\rangle\cdots
|e_{i_m}^{(m)}\rangle\right)\geq 0$ for all $i_j=1,\ldots,d_j$ and
$j=1,\ldots,m$. The $d_1\times\cdots\times d_m$ mutliway array
consisting of $a_{i_1\ldots i_m}$ is denoted by ${\cal A}_{\Psi}$.
When ${\mathcal{H}}_1=\cdots={\mathcal{H}}_m$, ${\cal A}_{\Psi}$ is
symmetric if and only if $|\Psi\rangle$ is symmetric in the sense of
quantum information \cite{hkwg09,hmmov09,nc00}. The geometric measure of symmetric states attracted much attention recently \cite{hmmov09,cxz10,odv08,hkwg09}.

When $|\Psi\rangle$ is symmetric, \reff{me} reduces to \cite{hkwg09}
\begin{eqnarray}\label{me-s}
G(\Psi)=\max_{|\Phi\rangle=|\phi\rangle^{\otimes m}}|\langle\Psi|\Phi\rangle|.
\end{eqnarray}

\subsection{Z-eigenvalues of a tensor (hypermatrix)}

For a tensor (or hypermatrix) ${\cal T}$ of order $m$ and dimension
$n$ with $m,n\geq 2$, we mean a multiway array consisting of numbers
$t_{i_1\cdots i_m}\in\mathbb{R}$ for all $i_j\in\{1,\ldots,n\}$ and
$j\in\{1,\ldots,m\}$. The set of all $m$-th order $n$ dimensional
tensors is denoted by
$\mathbb{R}^{m,n}$. Given a vector $\mathbf{x}\in \mathbb{C}^n$,
define ${\cal T}\mathbf{x}^{m-1}$ as an $n$-dimensional vector with
its $i$-th element being $\sum_{i_2,\ldots,i_m=1}^nt_{ii_2\cdots
i_m}\mathbf{x}_{i_2}\cdots\mathbf{x}_{i_m}$. Z-eigenvalues of
tensors were introduced by Qi \cite{q05}. Suppose that $\cal
T$ is a real tensor, i.e., ${\cal T}\in\mathbb{R}^{m,n}$. A number $\lambda\in\mathbb{R}$ is called a
Z-eigenvalue of ${\cal T}$, if it, together with a nonzero vector
$\mathbf{x}\in\mathbb{R}^n$, satisfies
\begin{eqnarray}\label{ze}
\left\{\begin{array}{ccc}{\cal T}\mathbf{x}^{m-1}&=&\lambda \mathbf{x},\\\mathbf{x}^T\mathbf{x}&=&1.\end{array}\right.
\end{eqnarray}
$\mathbf{x}$ is then called an associated Z-eigenvector of the
Z-eigenvalue $\lambda$, and $(\lambda,\mathbf{x})$ is called a
Z-eigenpair. Obviously, $\lambda={\cal
T}\mathbf{x}^m:=\sum_{i_1,\ldots,i_m=1}^nt_{i_1\cdots
i_m}\mathbf{x}_{i_1}\cdots\mathbf{x}_{i_m}$ for a Z-eigenpair
$(\lambda,\mathbf{x})$ of ${\cal T}$. A tensor ${\cal T}\in\mathbb{R}^{m,n}$
is called nonnegative, if $t_{i_1\cdots i_m}\geq 0$ for all
$i_j\in\{1,\ldots,n\}$ and $j\in\{1,\ldots,m\}$.

Many interesting results on Z-eigenvalues of tensors were obtained
very recently \cite{cs11,cpz11,nqww07,q05,q07,lqz12,l05},
especially, for nonnegative tensors. These results give insights on
the behaviors of the Z-eigenvalues and powerful numerical algorithms
for computing the Z-spectral radius of a nonnegative tensor, please
see \cite{cpz11} and references therein.

\section{A symmetric pure state with nonnegative amplitudes}

\setcounter{Assumption}{0}
\setcounter{Theorem}{0} \setcounter{Proposition}{0}
\setcounter{Corollary}{0} \setcounter{Lemma}{0}
\setcounter{Definition}{0} \setcounter{Remark}{0}
\setcounter{Algorithm}{0}  \setcounter{Example}{0}

\hspace{4mm} In this section, we establish a connection between the
geometric measure of entanglement for a symmetric pure state with nonnegative
amplitudes and the Z-spectral theory of nonnegative tensors. Based on this connection, the computation of the geometric measure for such states is investigated.

\subsection{A connection}
The following result was established in \cite[Theorem 1]{hmmov09}:
\begin{Proposition}\label{prop-1}
If $|\Psi\rangle\in{\mathcal{H}}$ is symmetric and nonnegative with the underlying orthonormal basis $\{|e_i\rangle\}_{i=1}^n$, then $|\Phi\rangle=|\phi\rangle^{\otimes m}$ in \reff{me-s} can be chosen with $\langle e_i|\phi\rangle\geq 0$ for all $i=1,\ldots,n$.
\end{Proposition}

Denote by $\mathbb{R}_+^n$ the nonnegative orthant of
$\mathbb{R}^n$, $\mathbb{R}_{++}^n$ the interior of $\mathbb{R}_+^n$, and ${\cal S}^{n-1}$ the unit sphere in
$\mathbb{R}^n$. Then, we have the following result.
\begin{Corollary}\label{cor-1}
If $|\Psi\rangle\in{\mathcal{H}}$ is symmetric and nonnegative, then
\begin{eqnarray}\label{gm-n}
G(\Psi)=\max_{\mathbf{x}\in \mathbb{R}_+^n\cap{\cal S}^{n-1}}{\mathcal{A}}_{\Psi}\mathbf{x}^m.
\end{eqnarray}
\end{Corollary}

\noindent {\bf Proof.} Under the orthonormal basis which makes ${\cal A}_{\Psi}$ nonnegative, we have that
\begin{eqnarray}\label{cor-1-1}
G(\Psi)=\max_{\mathbf{x}^{H}\mathbf{x}=1}|{\mathcal{A}}_{\Psi}\mathbf{x}^m|.
\end{eqnarray}
Here the superscript $^H$ means conjugate transpose. For any $\hat
\mathbf{x}$ with $\hat \mathbf{x}^H\hat \mathbf{x}=1$ being an
optimal solution for problem \reff{cor-1-1}, let $\mathbf{x}=|\hat
\mathbf{x}|$ be the componentwise module of $\hat \mathbf{x}$. Then,
$\mathbf{x}^T\mathbf{x}=1$ and
\begin{eqnarray*}
{\mathcal{A}}_{\Psi}\mathbf{x}^m\leq
|{\mathcal{A}}_{\Psi}\hat\mathbf{x}^m|\leq{\mathcal{A}}_{\Psi}|\hat\mathbf{x}|^m={\mathcal{A}}_{\Psi}\mathbf{x}^m.
\end{eqnarray*}
Here the first inequality follows from the facts that $\hat
\mathbf{x}$ is optimal and $\mathbf{x}$ is feasible for
\reff{cor-1-1}, and the second from the fact that ${\cal A}_{\Psi}$
is nonnegative. Consequently, the result \reff{gm-n} follows. \ep

To establish a connection, we present some basic results on the
Z-eigenvalues of nonnegative tensors. The following concept is
important for nonnegative tensors. ${\cal T}=(t_{i_1i_2\ldots i_m})$
is called {\em reducible} if there exists a nonempty proper index
subset $I\subset \{1,\ldots,n\}$ such that
\begin{eqnarray*}
t_{i_1i_2\ldots i_m}=0,\quad \forall i_1\in I,\quad \forall
i_2,\ldots,i_m\notin I.
\end{eqnarray*}
If ${\cal T}$ is not reducible, then ${\cal T}$ is called {\em irreducible}. Denote by ${\cal Z}({\cal T})$ the set of all Z-eigenvalues of tensor ${\cal T}$ and $\varrho({\cal T}):=\max\{|\lambda|\;|\;\lambda\in{\cal Z}({\cal T})\}$.

\begin{Proposition}\label{prop-2}
Let ${\cal T}\in\mathbb{R}^{m,n}$.
\begin{itemize}
\item [(a)] Every symmetric tensor ${\cal T}$ has at most $\frac{(m-1)^n-1}{m-2}$ Z-eigenvalues.
\item [(b)] If ${\cal T}$ is nonnegative, then there exists a nonnegative Z-eigenpair $(\lambda_0, \mathbf{x}^{(0)})$, i.e., $\lambda_0\geq 0$ and $\mathbf{x}^{(0)}\in \mathbb{R}^n_+\cap{\cal S}^{n-1}$. If ${\cal T}$ is furthermore irreducible, then $\lambda_0>0$ and $\mathbf{x}^{(0)}\in\mathbb{R}^n_{++}$.
\item [(c)] If ${\cal T}$ is nonnegative and symmetric, then $\varrho({\cal T})\in{\cal Z}({\cal T})$ and
    \begin{eqnarray*}
    \varrho({\cal T})=\max_{\mathbf{x}\in{\cal S}^{n-1}} {\cal T}\mathbf{x}^m=\max_{\mathbf{x}\in \mathbb{R}^n_+\cap {\cal S}^{n-1}} {\cal T}\mathbf{x}^m.
    \end{eqnarray*}
\end{itemize}
\end{Proposition}

\noindent {\bf Proof.} (a) follows from \cite[Theorem 5.6]{cs11}, (b) follows from \cite[Theorems 2.1 and 2.2]{cpz11}, and (c) follows from \cite[Theorem 3.10]{cpz11}. \ep

We call $\varrho({\cal T})$ the Z-spectral radius of tensor ${\cal
T}$ \cite{cpz11}. For nonnegative ${\cal T}\in\mathbb{R}^{m,n}$, let $\Lambda({\cal T})$ be the set of $\lambda\geq 0$ such that $\lambda$ together with some $\mathbf{x}\in\mathbb{R}^n_{+}\cap{\cal S}^{n-1}$ is a Z-eigenpair of ${\cal T}$. Then, $\varrho({\cal T})=\max\{\lambda\;|\;\lambda\in\Lambda({\cal T})\}$ by Proposition \ref{prop-2}.
Based on Corollary \ref{cor-1} and Proposition
\ref{prop-2}, we now establish the connection.
\begin{Theorem}\label{thm-1}
If $|\Psi\rangle\in{\mathcal{H}}$ is symmetric and nonnegative, then
\begin{eqnarray*}
G(\Psi)=\varrho({\cal A}_{\Psi}).
\end{eqnarray*}
\end{Theorem}

\subsection{Computation}

Theorem \ref{thm-1} shows that the geometric measure of symmetric pure
states with nonnegative amplitudes \cite{hmmov09,cxz10} can be computed through finding
the Z-spectral radii of the underlying nonnegative tensors. In this subsection, based on Theorem \ref{thm-1}, the computation of the geometric measure of such states is discussed.

{\bf 3.2.1. Multipartite qubit states.} For symmetric pure
multipartite qubit states with nonnegative amplitudes, \cite[Section
II.A]{cxz10} converts the geometric measure $G(\Psi)$ into a
polynomial rational fraction in one variable. By the derivatives,
$G(\Psi)$ can be computed. On the other hand, with \cite[Theorem
1]{qww09}, $G(\Psi)$ can be computed through finding the roots of an
univariate polynomial of degree $m$. These two methods are somewhat
different but both of them are the variable elimination methods.

{\bf 3.2.2. Multipartite qutrit states.} We consider multipartite
qutrit states in this part. The separability and measure of qutrit
entanglement were discussed by Caves and Mulburn \cite{cm99}, and
Hassan and Joag \cite{hj09}, and attracted much attention.

Based on the variable elimination method \cite[Theorem 3]{qww09} and \cite[Appendix]{qyw10}, we can compute the geometric measure of entanglement for symmetric pure multipartite qutrit states with nonnegative amplitudes.

In the sequel, we present the details. Let ${\cal H}_1=\cdots={\cal H}_m$, $d_1=\cdots=d_m=3$, and ${\cal H}:=\bigotimes_{k=1}^m{\cal H}_k$. Given a symmetric pure state $|\Psi\rangle\in{\cal H}$, if $|\Psi\rangle$ is nonnegative, i.e., there exists a basis $\{|e_i\rangle\}_{i=1}^3$ such that the tensor ${\cal A}_{\Psi}$ is nonnegative, then $G(\Psi)$ is equal to the Z-spectral radius of the tensor ${\cal A}_{\Psi}$ by Theorem \ref{thm-1}.

Denote by $a_{i_1i_2\ldots i_m}$ the $(i_1,i_2,\ldots,i_m)$-th element of the tensor ${\cal A}_{\Psi}$, we have the system of the Z-eigenvalue equations \reff{ze} of the tensor ${\cal A}_{\Psi}$ is:
\begin{eqnarray}\label{trze}
\left\{\begin{array}{ccl}\sum_{i_2,\ldots,i_m=1}^3a_{1i_2\ldots i_m}\mathbf{x}_{i_2}\cdots\mathbf{x}_{i_m}&=&\lambda\mathbf{x}_1,\\
\sum_{i_2,\ldots,i_m=1}^3a_{2i_2\ldots i_m}\mathbf{x}_{i_2}\cdots\mathbf{x}_{i_m}&=&\lambda\mathbf{x}_2,\\
\sum_{i_2,\ldots,i_m=1}^3a_{3i_2\ldots i_m}\mathbf{x}_{i_2}\cdots\mathbf{x}_{i_m}&=&\lambda\mathbf{x}_3,\\
\mathbf{x}_1^2+\mathbf{x}_2^2+\mathbf{x}_3^2&=&1.
\end{array}\right.
\end{eqnarray}

Note that $\lambda={\cal A}_{\Psi}\mathbf{x}^3$ for any Z-eigenpair $(\lambda,\mathbf{x})$ of ${\cal A}_{\Psi}$. We now give the following algorithm.
\begin{Algorithm}\label{algo2} (Symmetric multipartite qutrit states with nonnegative amplitudes)
{\rm \begin{description}
\item [Step 0] Input data ${\cal A}_{\Psi}$ with elements $a_{i_1i_2\ldots i_m}$. Set $\Pi({\cal A}_{\Psi})$ as the empty set.
\item [Step 1] If $a_{21\ldots1}=a_{31\ldots 1}=0$, then $\lambda=a_{11\ldots 1}$ is a Z-eigenvalue with a Z-eigenvector $\mathbf{x}:=(1,0,0)$. In this case, put $\lambda=a_{11\ldots 1}$ into $\Pi({\cal A}_{\Psi})$.
\item [Step 2] In this step, we consider Z-eigenvectors with $\mathbf{x}_3=0$ but $\mathbf{x}_2\neq 0$. If $\mathbf{x}_3=0$ and $\mathbf{x}_2\neq 0$, then \reff{trze} becomes:
\begin{eqnarray}\label{trze-1}
\left\{\begin{array}{ccl}\sum_{i_2,\ldots,i_m=1}^2a_{1i_2\ldots i_m}\mathbf{x}_{i_2}\cdots\mathbf{x}_{i_m}&=&\lambda\mathbf{x}_1,\\
\sum_{i_2,\ldots,i_m=1}^2a_{2i_2\ldots i_m}\mathbf{x}_{i_2}\cdots\mathbf{x}_{i_m}&=&\lambda\mathbf{x}_2,\\
\sum_{i_2,\ldots,i_m=1}^2a_{3i_2\ldots i_m}\mathbf{x}_{i_2}\cdots\mathbf{x}_{i_m}&=&0,\\
\mathbf{x}_1^2+\mathbf{x}_2^2&=&1.
\end{array}\right.
\end{eqnarray}
\begin{itemize}
\item [(i)] Multiply the first equation of \reff{trze-1} by $\mathbf{x}_2$ and the second by $\mathbf{x}_1$, then we get
\begin{eqnarray*}
\mathbf{x}_2\sum_{i_2,\ldots,i_m=1}^2a_{1i_2\ldots
i_m}\mathbf{x}_{i_2}\cdots\mathbf{x}_{i_m}=\mathbf{x}_1\sum_{i_2,\ldots,i_m=1}^2a_{2i_2\ldots
i_m}\mathbf{x}_{i_2}\cdots\mathbf{x}_{i_m}.
\end{eqnarray*}
Divide it by $\mathbf{x}_2^m$, the third equation of
\reff{trze-1} by $\mathbf{x}_2^{m-1}$, and set
$t=\frac{\mathbf{x}_1}{\mathbf{x}_2}$. Then, we get two polynomial
equations in variable $t$ as $f(t)=g(t)$ and $h(t)=0$.
\item [(ii)] For all nonnegative $t$ such that both $f(t)-g(t)=0$ and $h(t)=0$, we have $\mathbf{x}:=\left(\frac{t}{\sqrt{1+t^2}},\frac{1}{\sqrt{1+t^2}},0\right)$ is a Z-eigenvector of ${\cal A}_{\Psi}$. The corresponding Z-eigenvalue is $\lambda={\cal A}_{\Psi}\mathbf{x}^3$. Put these $\lambda$ into $\Pi({\cal A}_{\Psi})$.
\end{itemize}
\item [Step 3] In this step, we consider Z-eigenvectors with $\mathbf{x}_3\neq 0$. If $\mathbf{x}_3\neq 0$, then we have the following:
\begin{itemize}
\item [(i)] Multiply the first equation of \reff{trze} by $\mathbf{x}_3$ and the third by $\mathbf{x}_1$, then we get
\begin{eqnarray*}
\mathbf{x}_3\sum_{i_2,\ldots,i_m=1}^3a_{1i_2\ldots
i_m}\mathbf{x}_{i_2}\cdots\mathbf{x}_{i_m}=\mathbf{x}_1\sum_{i_2,\ldots,i_m=1}^3a_{3i_2\ldots
i_m}\mathbf{x}_{i_2}\cdots\mathbf{x}_{i_m}.
\end{eqnarray*}
Similarly, we have
\begin{eqnarray*}
\mathbf{x}_3\sum_{i_2,\ldots,i_m=1}^3a_{2i_2\ldots
i_m}\mathbf{x}_{i_2}\cdots\mathbf{x}_{i_m}=\mathbf{x}_2\sum_{i_3,\ldots,i_m=1}^3a_{3i_2\ldots
i_m}\mathbf{x}_{i_2}\cdots\mathbf{x}_{i_m}.
\end{eqnarray*}

Divide them by $\mathbf{x}_3^m$ respectively, and set $u=\frac{\mathbf{x}_1}{\mathbf{x}_3}$ and
$v=\frac{\mathbf{x}_2}{\mathbf{x}_3}$. Then, we get two polynomial
equations in variables $u$ and $v$ as $f(u,v)=0$ and $g(u,v)=0$.
\item [(ii)] Write them in univariate polynomial equations in the variable $u$ with coefficients as univariate polynomials in the variable $v$ as
    \begin{eqnarray}\label{trze-2}
    \begin{array}{ccl}
    f(u,v)&=&a_0(v)u^{m}+a_1(v)u^{m-1}+\cdots+a_{m}(v)=0,\\
    g(u,v)&=&b_0(v)u^{m-1}+b_1(v)u^{m-2}+\cdots+b_{m-1}(v)=0.
    \end{array}
    \end{eqnarray}
\item [(iii)] Form the $(2m-1)\times (2m-1)$ polynomial matrix (the Sylvester matrix \cite{cls98}) in the variable $v$ as:
    \begin{eqnarray*}
    M(v):=\left(\begin{array}{cccccccc}a_0(v)&a_1(v)&\cdots&a_{m}(v)&0&0&\cdots&0\\0&a_0(v)&a_1(v)&\cdots&a_{m}(v)&0&\cdots&0\\
    &\ddots&\ddots&\ddots&\ddots&&&\\0&0&0&0&a_0(v)&a_1(v)&\cdots&a_{m}(v)\\
    b_0(v)&b_1(v)&\cdots&b_{m-1}(v)&0&0&\cdots&0\\0&b_0(v)&b_1(v)&\cdots&b_{m-1}(v)&0&\cdots&0\\
    &\ddots&\ddots&\ddots&\ddots&&&\\0&0&0&0&b_0(v)&b_1(v)&\cdots&b_{m-1}(v)
    \end{array}\right).
    \end{eqnarray*}
\item [(iv)] Compute the determinant of the matrix $M(v)$ as a univariate polynomial in the variable $v$. Denote it by $d(v)$.
\item [(v)] For any nonnegative $v$ such that $d(v)=0$ and then any nonnegative $u$ such that $(u,v)$ being a solution for \reff{trze-2},  $\mathbf{x}:=\left(\frac{u}{\sqrt{1+u^2+v^2}},\frac{v}{\sqrt{1+u^2+v^2}},\frac{1}{\sqrt{1+u^2+v^2}}\right)$ is a Z-eigenvector of ${\cal A}_{\Psi}$. The corresponding Z-eigenvalue is $\lambda={\cal A}_{\Psi}\mathbf{x}^3$. Put all such $\lambda$ into $\Pi({\cal A}_{\Psi})$.
\end{itemize}
\end{description}}
\end{Algorithm}

\begin{Theorem}\label{thm-3}
If $|\Psi\rangle\in{\mathcal{H}}=\bigotimes_{k=1}^m{\cal H}_k$ is a symmetric nonnegative qutrit state and $\Pi({\cal A}_{\Psi})$ is generated by Algorithm \ref{algo2}, then $\Pi({\cal A}_{\Psi})=\Lambda({\cal A}_{\Psi})$, and
\begin{eqnarray*}
G(\Psi)=\varrho({\cal A}_{\Psi})=\max\{\lambda\;|\;\lambda\in\Pi({\cal A}_{\Psi})\}.
\end{eqnarray*}
\end{Theorem}

\noindent {\bf Proof.} By Steps 1-3 of Algorithm \ref{algo2} and the Sylvester theorem \cite{cls98} used in Step 3, all the nonnegative Z-eigenvectors are considered. Consequently, $\Pi({\cal A}_{\Psi})=\Lambda({\cal A}_{\Psi})$. Now, by Proposition \ref{prop-2} and Theorem \ref{thm-1}, the result follows. \ep

This result is new for computing $G(\Psi)$ in the qutrit case.

{\bf 3.2.3. Z-spectra of multipartite qubit and qutrit states.}
In the following, we show the Z-spectra for GHZ, $W$, and inverted-$W$ states, and a qutrit state. As in \cite{wg03}, define
\begin{eqnarray*}
|S(m,k)\rangle:=\sqrt{\frac{k!(m-k)!}{m!}}\sum_{\tau\in\mathfrak{G}_m}|\tau(\underbrace{0\cdots 0}_{k}\underbrace{1\cdots 1}_{m-k})\rangle,
\end{eqnarray*}
here $\mathfrak{G}_m$ is the symmetric group on $m$ elements. Examples \ref{exm-1}, \ref{exm-2} and \ref{exm-3} are computed through Theorem 1 in \cite{qww09}.
\begin{Example}\label{exm-1}
{\rm
The $m$GHZ state is defined as:
\begin{eqnarray*}
|mGHZ\rangle:=\left(|S(m,0)\rangle+|S(m,m)\rangle\right)/\sqrt{2}.
\end{eqnarray*}
Under the basis $\{|0\rangle,|1\rangle\}$, we have ${\cal
A}_{mGHZ}\in\mathbb{R}^{m,2}$ and the Z-eigenpairs are:
\begin{eqnarray*}
\begin{array}{c}
\left(\frac{1}{\sqrt{2}},(1,0)\right),
\left(\frac{1}{\sqrt{2}},(0,1)\right), \;\mbox{and}\;
\left(\frac{1}{\sqrt{2}^{m-1}},\left(\frac{1}{\sqrt{2}},\frac{1}{\sqrt{2}}\right)\right)
\end{array}
\end{eqnarray*}
and five more when $m$ is odd:
\begin{eqnarray*}
\begin{array}{c}
\left(\frac{1}{\sqrt{2}},(-1,0)\right), \left(\frac{1}{\sqrt{2}},(0,-1)\right),\left(\frac{1}{\sqrt{2}^{m-1}},\left(-\frac{1}{\sqrt{2}},\frac{1}{\sqrt{2}}\right)\right),\\
\left(\frac{1}{\sqrt{2}^{m-1}},\left(\frac{1}{\sqrt{2}},-\frac{1}{\sqrt{2}}\right)\right),
\;\mbox{and}\;
\left(\frac{1}{\sqrt{2}^{m-1}},\left(-\frac{1}{\sqrt{2}},-\frac{1}{\sqrt{2}}\right)\right).
\end{array}
\end{eqnarray*}
We have, $G(mGHZ)=\varrho({\cal A}_{mGHZ})=\frac{1}{\sqrt{2}}$ which agrees with that in \cite[Section II.A]{wg03}, and it can be attained with nonnegative Z-eigenvectors which agrees with Proposition \ref{prop-1}. The corresponding nearest separable state is $|\Phi\rangle=|\phi\rangle^{\otimes m}$ with $|\phi\rangle:=|0\rangle$ or $|1\rangle$.}
\end{Example}

\begin{Example}\label{exm-2}
{\rm
In this example, $W$ state for a $3$-partite qubit setting is considered. The $W$ state is defined as:
\begin{eqnarray*}
|W\rangle:=|S(3,2)\rangle=\left(|001\rangle+|010\rangle+|100\rangle\right)/\sqrt{3}.
\end{eqnarray*}
Under the basis $\{|0\rangle,|1\rangle\}$, we have ${\cal A}_{W}\in\mathbb{R}^{3,2}$ and the Z-eigenpairs are:
\begin{eqnarray*}
\begin{array}{c}
\left(0,(0,1)\right),\left(\frac{2}{3},\left(\sqrt{\frac{2}{3}},\sqrt{\frac{1}{3}}\right)\right),\left(\frac{2}{3},\left(-\sqrt{\frac{2}{3}},\sqrt{\frac{1}{3}}\right)\right),\\
\left(-\frac{2}{3},\left(\sqrt{\frac{2}{3}},-\sqrt{\frac{1}{3}}\right)\right),
\;\mbox{and}\;
\left(-\frac{2}{3},\left(-\sqrt{\frac{2}{3}},-\sqrt{\frac{1}{3}}\right)\right).
\end{array}
\end{eqnarray*}
Again, $G({\cal A}_{W})=\varrho({\cal A}_{W})=\frac{2}{3}$ which agrees with that in \cite[Section II.A]{wg03}, and it can be attained with nonnegative Z-eigenvectors which agrees with Proposition \ref{prop-1}. The corresponding nearest separable state is $|\Phi\rangle=|\phi\rangle^{\otimes 3}$ with $|\phi\rangle:=\sqrt{\frac{2}{3}}|0\rangle+\sqrt{\frac{1}{3}}|1\rangle$.}
\end{Example}

\begin{Example}\label{exm-3}
{\rm In this example, inverted-$W$ state for a $3$-partite qubit setting is considered. It is defined as
\begin{eqnarray*}
|\widetilde{W}\rangle:=|S(3,1)\rangle=\left(|110\rangle+|101\rangle+|011\rangle\right)/\sqrt{3}.
\end{eqnarray*}
Similarly, we have ${\cal A}_{\widetilde{W}}\in\mathbb{R}^{3,2}$.
After switching $\mathbf{x}_1$ and $\mathbf{x}_2$, the Z-eigenvalues equations \reff{ze} of $\widetilde{W}$ becomes that for $W$. Consequently, $G(\widetilde{W})=\varrho({\cal A}_{\widetilde{W}})=\frac{2}{3}$ by Example \ref{exm-2} with the corresponding nearest separable state being $|\Phi\rangle=|\phi\rangle^{\otimes 3}$ with $|\phi\rangle:=\sqrt{\frac{1}{3}}|0\rangle+\sqrt{\frac{2}{3}}|1\rangle$.}
\end{Example}

\begin{Example}\label{exm-4}
{\rm In this example, a general GHZ state \cite[Eq. (9)]{hj09} for a $3$-partite qutrit setting is considered. It is defined as
\begin{eqnarray*}
|\Psi\rangle:=\alpha|111\rangle+\beta|222\rangle+\gamma|333\rangle,\;\alpha^2+\beta^2+\gamma^2=1.
\end{eqnarray*}
Here $\{|1\rangle,|2\rangle,|3\rangle\}$ is the basis for each qutrit. We see that ${\cal A}_{\Psi}\in\mathbb{R}^{3,3}$ is nonnegative and symmetric when $\alpha,\beta,\gamma\geq 0$. In this situation, the Z-eigenvalue equations \reff{trze} become:
\begin{eqnarray*}
\alpha \mathbf{x}_1^2=\lambda\mathbf{x}_1,\;\beta \mathbf{x}_2^2=\lambda\mathbf{x}_2,\;\gamma \mathbf{x}_3^2=\lambda\mathbf{x}_3,\;\mbox{and}\;\mathbf{x}_1^2+\mathbf{x}_2^2+\mathbf{x}_3^2=1.
\end{eqnarray*}
When $\alpha\beta\gamma=0$, the Z-spectra can be computed through Theorem 1 in \cite{qww09}. We mainly consider the nondegenerate case when $\alpha\beta\gamma>0$.
By Algorithm \ref{algo2}, we can compute all the nonnegative Z-eigenpairs as:
\begin{eqnarray*}
\begin{array}{c}
(\alpha,(1,0,0)),\;(\beta,(0,1,0)),\;(\gamma,(0,0,1)),\;\left(\frac{\alpha\beta}{\sqrt{\alpha^2+\beta^2}},
(\frac{\beta}{\sqrt{\alpha^2+\beta^2}},\frac{\alpha}{\sqrt{\alpha^2+\beta^2}},0)\right),\\
\left(\frac{\alpha\gamma}{\sqrt{\alpha^2+\gamma^2}},
(\frac{\gamma}{\sqrt{\alpha^2+\gamma^2}},0,\frac{\alpha}{\sqrt{\alpha^2+\gamma^2}})\right),\;\left(\frac{\beta\gamma}{\sqrt{\beta^2+\gamma^2}},
(0,\frac{\gamma}{\sqrt{\beta^2+\gamma^2}},\frac{\beta}{\sqrt{\beta^2+\gamma^2}})\right),\;\mbox{and}\\
\left(\frac{\alpha\beta\gamma}{\tau},
(\frac{\beta\gamma}{\tau},\frac{\alpha\gamma}{\tau},
\frac{\alpha\beta}{\tau})\right)\;\mbox{with}\;\tau:=\sqrt{\alpha^2\beta^2+\beta^2\gamma^2+\alpha^2\gamma^2}.
\end{array}
\end{eqnarray*}
We see that $G(\Psi)=\varrho({\cal A}_{\Psi})=\max\{\alpha,\beta,\gamma\}$. The corresponding nearest separable state is $|\Phi\rangle=|\phi\rangle^{\otimes 3}$ with $|\phi\rangle:=|1\rangle$ when $G(\Psi)=\alpha$, $|2\rangle$ when $G(\Psi)=\beta$ and $|3\rangle$ when $G(\Psi)=\gamma$. }
\end{Example}

{\bf 3.2.4. Multipartite qudit states.}
For symmetric pure multipartite qudit states, there is no analogue result of Proposition \ref{prop-2} (a) for the quantum eigenvalue problem \reff{qe}. Consequently, the computation for the largest quantum eigenvalues of a state is very complicated in general.

Nonetheless, for symmetric pure states $|\Psi\rangle$ with
nonnegative irreducible ${\cal A}_{\Psi}$, Proposition \ref{prop-2}
(a) helps to prove that the shifted higher order power method
\cite{km11} is locally convergent, i.e., with the initial point
sufficiently close to a Z-eigenvector of the Z-spectral radius, the
shifted higher order power method converges to the Z-spectral radius
(Lemma \ref{lem-1-a}). Consequently, if we randomly choose
initial points in $\mathbb{R}^n_{++}\cap{\cal S}^{n-1}$, then with a
positive probability we can find the Z-spectral radius for an
irreducible nonnegative tensor (Theorem \ref{thm-1-a}). This,
together with Theorem \ref{thm-1}, implies that Algorithm \ref{algo}
can find the geometric measure of symmetric pure states
$|\Psi\rangle$ with nonnegative irreducible ${\cal A}_{\Psi}$. We
present the details in Appendix. When ${\cal A}_{\Psi}$ is
reducible, the technique in \cite{hhq10} is applicable.

Besides the shifted higher order power method, a generalized Newton method like that in \cite{lqy12} may be developed for finding the Z-spectral radius.

\section{A nonsymmetric pure state with nonnegative amplitudes}

\setcounter{Assumption}{0}
\setcounter{Theorem}{0} \setcounter{Proposition}{0}
\setcounter{Corollary}{0} \setcounter{Lemma}{0}
\setcounter{Definition}{0} \setcounter{Remark}{0}
\setcounter{Algorithm}{0}  \setcounter{Example}{0}

\hspace{4mm} In this section, we extend the results in the last section to nonsymmetric pure states with nonnegative amplitudes. To this end, we need the spectral theory for multilinear forms \cite{l05,fgh11,rv11}. We first establish analogue results for nonnegative multilinear forms.

Let ${\cal A}=(a_{i_1\ldots i_m})$ be a $d_1\times\cdots\times d_m$ real tensor (hypermatrix).  $\sigma\in\mathbb{R}$ is called a singular value of ${\cal A}$, if it, together with $\mathbf{x}^{(1)}\in\mathbb{R}^{d_1}\cap{\cal
S}^{d_1-1}$, $\ldots$, $\mathbf{x}^{(m)}\in\mathbb{R}^{d_m}\cap{\cal
S}^{d_m-1}$, satisfies
\begin{eqnarray}\label{sing}
\sum_{1\leq i_j\leq d_j,\;j\neq k}a_{i_1\ldots
i_m}\mathbf{x}_{i_1}^{(1)}\cdots\mathbf{x}_{i_m}^{(m)}=\sigma \mathbf{x}^{(k)}_{i_k},\;\forall i_k=1,\ldots,d_k,\;\forall k=1,\ldots,m.
\end{eqnarray}
The vector $\mathbf{x}^{(k)}$ is called the mode-$k$ singular vector corresponding to the singular value $\sigma$ \cite{l05,rv11}. Denote the largest singular value of $\cal A$ by $\sigma({\cal A})$.

\begin{Proposition}\label{prop-5}
Let ${\cal A}=(a_{i_1\ldots i_m})$ be a $d_1\times\cdots\times d_m$ real tensor. Then,
\begin{eqnarray}\label{sin-1}
\sigma({\cal A})=\max_{\mathbf{x}^{(1)}\in {\cal
S}^{d_1-1},\ldots,\mathbf{x}^{(m)}\in {\cal
S}^{d_m-1}}{\mathcal{A}}\mathbf{x}^{(1)}\cdots
\mathbf{x}^{(m)}:=\sum_{i_1=1}^{d_1}\cdots\sum_{i_m=1}^{d_m}a_{i_1\ldots
i_m}\mathbf{x}_{i_1}^{(1)}\cdots\mathbf{x}_{i_m}^{(m)}.
\end{eqnarray}
Moreover, if ${\cal A}$ is nonnegative, then
the mode-$k$ singular vector corresponding to $\sigma({\cal A})$ can be chosen nonnegative.
\end{Proposition}

\noindent {\bf Proof.} We see firstly from \reff{sing} that for any singular value $\sigma$ of ${\cal A}$ with singular vectors $\mathbf{x}^{(1)},\ldots,\mathbf{x}^{(m)}$, we have $\sigma={\mathcal{A}}\mathbf{x}^{(1)}\cdots
\mathbf{x}^{(m)}$. Secondly, by optimization theory, the singular vectors are exactly the critical points of the maximization problem \reff{sin-1}. Hence, \reff{sin-1} follows.

Now, we show the second result. Suppose now that ${\cal A}$ is nonnegative.
We have
\begin{eqnarray*}
&&\max_{\mathbf{x}^{(1)}\in \mathbb{R}_+^{d_1}\cap{\cal
S}^{d_1-1},\ldots,\mathbf{x}^{(m)}\in \mathbb{R}_+^{d_m}\cap{\cal
S}^{d_m-1}}{\mathcal{A}}\mathbf{x}^{(1)}\cdots
\mathbf{x}^{(m)}\\
&\leq& \max_{\mathbf{x}^{(1)}\in {\cal
S}^{d_1-1},\ldots,\mathbf{x}^{(m)}\in {\cal
S}^{d_m-1}}{\mathcal{A}}\mathbf{x}^{(1)}\cdots
\mathbf{x}^{(m)}\\
&\leq&\max_{\mathbf{x}^{(1)}\in \mathbb{R}_+^{d_1}\cap{\cal
S}^{d_1-1},\ldots,\mathbf{x}^{(m)}\in \mathbb{R}_+^{d_m}\cap{\cal
S}^{d_m-1}}{\mathcal{A}}\mathbf{x}^{(1)}\cdots
\mathbf{x}^{(m)}.
\end{eqnarray*}
Here the second inequality follows from the fact that ${\cal A}$ is nonnegative.

Consequently,
\begin{eqnarray*}
\sigma({\cal A})=\max_{\mathbf{x}^{(1)}\in {\cal
S}^{d_1-1},\ldots,\mathbf{x}^{(m)}\in {\cal
S}^{d_m-1}}{\mathcal{A}}\mathbf{x}^{(1)}\cdots
\mathbf{x}^{(m)}=\max_{\mathbf{x}^{(1)}\in \mathbb{R}_+^{d_1}\cap{\cal
S}^{d_1-1},\ldots,\mathbf{x}^{(m)}\in \mathbb{R}_+^{d_m}\cap{\cal
S}^{d_m-1}}{\mathcal{A}}\mathbf{x}^{(1)}\cdots
\mathbf{x}^{(m)}.
\end{eqnarray*}
Hence, the optimal value $\sigma({\cal A})$ of \reff{sin-1} can be achieved with nonnegative $\mathbf{x}^{(k)}$'s. Then, there exist nonnegative $\mathbf{x}^{(k)}$'s that are critical points of \reff{sin-1}. By the correspondence of the critical points of \reff{sin-1} and the singular vectors of ${\cal A}$, the result follows. \ep

We now generalize Corollary \ref{cor-1} to a nonsymmetric pure state with
nonnegative amplitudes.
\begin{Proposition}\label{prop-3}
If $|\Psi\rangle\in{\mathcal{H}}$ is nonnegative with the underlying orthonormal bases $\{|e_i^{(k)}\rangle\}_{i=1}^{d_k}$ for $k=1,\ldots,m$, then $|\Phi\rangle=\bigotimes_{k=1}^m|\phi^{(k)}\rangle$ in \reff{me} can be chosen with $\langle e_i^{(k)}|\phi^{(k)}\rangle\geq 0$ for all $i=1,\ldots,d_k$ and $k=1,\ldots,m$. Consequently,
\begin{eqnarray*}
G(\Psi)=\max_{\mathbf{x}^{(1)}\in \mathbb{R}_+^{d_1}\cap{\cal
S}^{d_1-1},\ldots,\mathbf{x}^{(m)}\in \mathbb{R}_+^{d_m}\cap{\cal
S}^{d_m-1}}{\mathcal{A}}_{\Psi}\mathbf{x}^{(1)}\cdots
\mathbf{x}^{(m)}.
\end{eqnarray*}
\end{Proposition}

\noindent {\bf Proof.} It is similar to that for Corollary \ref{cor-1}. \ep

By Propositions \ref{prop-5} and \ref{prop-3}, we have the following theorem.
\begin{Theorem}\label{thm-2}
If $|\Psi\rangle\in{\mathcal{H}}$ is nonnegative with the underlying orthonormal bases $\{|e_i^{(k)}\rangle\}_{i=1}^{d_k}$ for $k=1,\ldots,m$, then
\begin{eqnarray*}
G(\Psi)=\sigma({\cal A}_\Psi).
\end{eqnarray*}
\end{Theorem}

Theorem \ref{thm-2} gives a connection between the geometric measure of nonsymmetric pure states with nonnegative amplitudes and the spectral theory of nonnegative tensors.

Proposition \ref{prop-5} generalizes Proposition \ref{prop-2} (b) and (c) to the context of nonnegative multilinear forms.
As it can be seen from the last section that Proposition \ref{prop-2} (a) plays an important role in the computational issues, we establish an analogue result for multilinear forms in the following. To this end, symmetric embedding introduced in \cite{rv11} is needed.

Let ${\cal A}=(a_{i_1\ldots i_m})$ be a $d_1\times\cdots\times d_m$ real tensor and ${\cal S}_{\cal A}$ be the symmetric embedding of tensor ${\cal A}$ \cite[Section 2.2 ]{rv11}. ${\cal S}_{\cal A}$ is an $m$-th order $\sum_{k=1}^md_k$ dimensional symmetric tensor.
Then, we have the following result.
\begin{Proposition}\label{prop-4}
Let ${\cal A}=(a_{i_1\ldots i_m})$ be a $d_1\times\cdots\times d_m$ real tensor. Then, $\sigma$ is a nonzero singular value of ${\cal A}$ if and only if $\frac{m!}{\sqrt{m^m}}\sigma$ is a nonzero Z-eigenvalue of ${\cal S}_{\cal A}$.
\end{Proposition}

\noindent {\bf Proof.} The ``only if" part follows from \cite[Theorem 4.7]{rv11}.

We show the ``if" part in the following. Now, suppose that $\mathbf{y}:=({\mathbf{y}^{(1)}}^T,\ldots,{\mathbf{y}^{(m)}}^T)^T\in\mathbb{R}^{d_1+\cdots+d_m}\cap{\cal
S}^{d_1+\cdots+d_m-1}$ with $\mathbf{y}^{(k)}\in\mathbb{R}^{d_k}$ for each $k$ is a Z-eigenvector of ${\cal S}_{\cal A}$ corresponding to Z-eigenvalue $\lambda\neq 0$. Suppose, without loss of generality, that $\mathbf{y}^{(1)}\neq\mathbf{0}$. By the definition of ${\cal S}_{\cal A}$, we have
\begin{eqnarray*}
\lambda(\mathbf{y}^{(1)})^T\mathbf{y}^{(1)}&=&\sum_{i_1=1}^{d_1}\mathbf{y}^{(1)}_{i_1}\left[\sum_{i_2,\ldots,i_m=1}^{d_1+\cdots+d_m}({\cal S}_{\cal A})_{i_1i_2\ldots i_m}\mathbf{y}_{i_2}\cdots\mathbf{y}_{i_m}\right]\\
&=&(m-1)!{\mathcal{A}}\mathbf{y}^{(1)}\cdots\mathbf{y}^{(m)}\\
&=&\sum_{i_k=1}^{d_k}\mathbf{y}^{(k)}_{i_k}\left[\sum_{1\leq i_j\leq d_1+\cdots+d_m,\;j\neq k}({\cal S}_{\cal A})_{i_1i_2\ldots i_m}\mathbf{y}_{i_1}\cdots\mathbf{y}_{i_m}\right]\\
&=&\lambda(\mathbf{y}^{(k)})^T\mathbf{y}^{(k)}
\end{eqnarray*}
for all $k=2,\ldots,m$. Consequently, $(\mathbf{y}^{(k)})^T\mathbf{y}^{(k)}=\frac{1}{m}$ for all $k=1,\ldots,m$. Moreover,
\begin{eqnarray*}
\sum_{i_2,\ldots,i_m=1}^{d_1+\cdots+d_m}({\cal S}_{\cal A})_{i_1i_2\ldots i_m}\mathbf{y}_{i_2}\cdots\mathbf{y}_{i_m}&=&(m-1)!\sum_{i_2=1}^{d_2}\cdots\sum_{i_m=1}^{d_m}a_{i_1i_2\ldots
i_m}\mathbf{y}_{i_2}^{(2)}\cdots\mathbf{y}_{i_m}^{(m)}\\
&=&\lambda \mathbf{y}^{(1)}_{i_1},\;\forall i_1=1,\ldots,d_1.
\end{eqnarray*}
Let $\mathbf{x}^{(k)}:=\sqrt{m}\mathbf{y}^{(k)}$ for all $k=1,\ldots,m$. We then have
\begin{eqnarray*}
(m-1)!\frac{1}{\sqrt{m^{m-1}}}\sum_{i_2=1}^{d_2}\cdots\sum_{i_m=1}^{d_m}a_{i_1i_2\ldots
i_m}\mathbf{x}_{i_2}^{(2)}\cdots\mathbf{x}_{i_m}^{(m)}=\lambda\frac{1}{\sqrt{m}} \mathbf{x}^{(1)}_{i_1},\;\forall i_1=1,\ldots,d_1.
\end{eqnarray*}
Similarly, we have
\begin{eqnarray*}
(m-1)!\frac{1}{\sqrt{m^{m-1}}}\sum_{1\leq i_j\leq d_j,\;j\neq k}a_{i_1\ldots
i_m}\mathbf{x}_{i_1}^{(1)}\cdots\mathbf{x}_{i_m}^{(m)}=\lambda\frac{1}{\sqrt{m}} \mathbf{x}^{(k)}_{i_k},\;\forall i_k=1,\ldots,d_k,\;\forall k=2,\ldots,m.
\end{eqnarray*}
This, together with \reff{sing}, implies that $\frac{\sqrt{m^m}}{m!}\lambda$ is a nonzero singular value of ${\cal A}$. The proof is complete.  \ep

\begin{Corollary}\label{cor-2}
Let ${\cal A}=(a_{i_1\ldots i_m})$ be a $d_1\times\cdots\times d_m$ real tensor. Then, it has at most $\frac{(m-1)^N-1}{m-2}$ singular values, here $N:=\sum_{k=1}^md_k$.
\end{Corollary}

\noindent {\bf Proof.} From the definition of ${\cal S}_{\cal A}$, it is easy to see that if $0$ is a singular value of tensor ${\cal A}$ with singular vectors $\mathbf{x}^{(1)},\ldots,\mathbf{x}^{(m)}$, then $0$ is a Z-eigenvalue of ${\cal S}_{\cal A}$ with Z-eigenvector $\mathbf{x}:=\left(\frac{\left(\mathbf{x}^{(1)}\right)^T}{\sqrt{m}},\ldots,\frac{\left(\mathbf{x}^{(m)}\right)^T}{\sqrt{m}}\right)^T$.

Now, the result follows from Proposition \ref{prop-2} (a) and Proposition \ref{prop-4} immediately. \ep

\begin{Corollary}\label{cor-3}
If $|\Psi\rangle\in{\mathcal{H}}$ is nonnegative with the underlying orthonormal bases $\{|e_i^{(k)}\rangle\}_{i=1}^{d_k}$ for $k=1,\ldots,m$, then
\begin{eqnarray*}
G(\Psi)=\sigma({\cal A}_\Psi)=\frac{\sqrt{m^m}}{m!}\varrho({\cal S}_{{\cal A}_{\Psi}}).
\end{eqnarray*}
\end{Corollary}

\noindent {\bf Proof.} It follows from Propositions \ref{prop-2} and \ref{prop-4}, and Theorem \ref{thm-2} immediately. \ep

So, for the geometric measure of
entanglement for pure states with nonnegative
amplitudes, the nonsymmetric ones can be converted to the symmetric ones. Consequently, the numerical methods in the last section are applicable.
\section{Conclusion}

\setcounter{Assumption}{0}
\setcounter{Theorem}{0} \setcounter{Proposition}{0}
\setcounter{Corollary}{0} \setcounter{Lemma}{0}
\setcounter{Definition}{0} \setcounter{Remark}{0}
\setcounter{Algorithm}{0}  \setcounter{Example}{0}

\hspace{4mm} We have established a connection between the geometric
measure of entanglement for a pure state with nonnegative amplitudes
and the spectral theory of nonnegative tensors. Especially, we have
shown that the geometric measure of entanglement of a symmetric pure
state with nonnegative amplitudes is equal to the Z-spectral radius
of the underlying nonnegative symmetric tensor, and the geometric
measure of entanglement of a nonsymmetric pure state with
nonnegative amplitudes is equal to the largest singular value of the
underlying nonnegative tensor. Based on this connection, the
computation of the geometric measure is investigated in
details. An analytical derivation for the geometric measure of
symmetric pure multipartite qutrit states is given. For multipartite
qudit states, an algorithm with randomization is proposed and its convergence is proved. We have proven that it is convergent. Independently,
results in Section 4 for nonnegative multilinear forms have their own
significance.

Many established results and algorithms for eigenvalues of
nonnegative tensors, e.g. the generalized Perron-Frobenius theorem
\cite{cpz08,cpz11,yy11,fgh11,hhq10}, can be applied to the study and
computation of the geometric measure of a pure state with
nonnegative amplitudes.
Also, there is a possibility that new ideas will emerge from the
intersection of the geometric measure theory of quantum entanglement
and the spectral theory of nonnegative tensors.


\appendix
\section{Appendix: An algorithm for finding the Z-spectral radius of a nonnegative irreducible symmetric tensor}

\setcounter{Assumption}{0}
\setcounter{Theorem}{0} \setcounter{Proposition}{0}
\setcounter{Corollary}{0} \setcounter{Lemma}{0}
\setcounter{Definition}{0} \setcounter{Remark}{0}
\setcounter{Algorithm}{0}  \setcounter{Example}{0}

In this section, we propose an algorithm for finding the Z-spectral radius of a nonnegative irreducible symmetric tensor, and prove its convergence. By Theorem \ref{thm-1}, a method for computing the geometric measure for symmetric pure states with nonnegative amplitudes is then given.
\subsection{The power method}
The following is the shifted higher order power method investigated in \cite{km11}.
\begin{Algorithm}\label{algo1} (Shifted Higher Order Power Method (SHOPM))
\begin{description}
\item [Step 0] Initialization: choose $\mathbf{x}^{(0)}\in \mathbb{R}^n_{++}\cap {\cal S}^{n-1}$ and $\alpha>0$.  Let $k:=0$ and $\lambda_0:={\cal A}(\mathbf{x}^{(0)})^m$.

\item [Step 1] Compute
\begin{eqnarray*}
\begin{array}{c}
\hat \mathbf{x}^{(k+1)}:={\cal A}(\mathbf{x}^{(k)})^{m-1}+\alpha \mathbf{x}^{(k)},\quad \mathbf{x}^{(k+1)}:=\frac{\hat \mathbf{x}^{(k+1)}}{\|\hat \mathbf{x}^{(k+1)}\|}, \quad \mbox{and}\quad \lambda_{k+1}:={\cal A}(\mathbf{x}^{(k+1)})^m.
\end{array}
\end{eqnarray*}

\item [Step 2]If ${\cal A}(\mathbf{x}^{(k+1)})^{m-1}=\lambda_{k+1}\mathbf{x}^{(k+1)}$, stop. Otherwise, let $k:=k+1$, go to Step 1.
\end{description}
\end{Algorithm}

By \cite[Theorem 4.4]{km11}, Proposition \ref{prop-2} and \cite{hkwg09}, we have the following result.
\begin{Proposition}\label{prop-3-a}
Let ${\cal A}\in\mathbb{R}^{m,n}$ be nonnegative and symmetric. For $\alpha>(m-1)\varrho({\cal A})$, the iterates $(\lambda_k,\mathbf{x}^{(k)})$ generated by Algorithm \ref{algo1} satisfy the following
properties.
\begin{itemize}
\item [(a)] The sequence ${\lambda_k}$ is nondecreasing and converges to a Z-eigenvalue $\lambda_*\geq 0$.
\item [(b)] If there are only finitely many Z-eigenvectors, the sequence $\{\mathbf{x}^{(k)}\}$ converges to a Z-eigenvector of ${\cal A}$ corresponding to $\lambda_*$.
\item [(c)] Sequence $\{\mathbf{x}^{(k)}\}$ has an accumulation point and every such point is a Z-eigenvector of ${\cal A}$ corresponding to $\lambda_*$.
\end{itemize}
\end{Proposition}

Based on Algorithm \ref{algo1}, we introduce an algorithm for finding the Z-spectral radius of a nonnegative irreducible symmetric tensor.
\begin{Algorithm}\label{algo} (An algorithm for the Z-spectral radius of a nonnegative irreducible symmetric tensor)
\begin{description}
\item [Step 0] Let $k:=0$ and compute the Z-eigenvalue $\lambda_0$ of ${\cal A}$ through Algorithm \ref{algo1}. Let ${\cal A}:=\frac{{\cal A}}{\lambda_0}$.

\item [Step 1] Choose $N$ initial points in $\mathbb{R}^n_{++}\cap{\cal S}^{n-1}$. For the $i$-th initial point, compute the Z-eigenvalue $\mu_i$ of ${\cal A}$ through Algorithm \ref{algo1}. Let $\mu:=\max_{1\leq i\leq N}\mu_i$.

\item [Step 2]If $\mu=1$, then $\lambda:=\prod_{0\leq j\leq k}\lambda_j$. The algorithm is terminated. Otherwise, let $\lambda_{k+1}:=\mu$, ${\cal A}:=\frac{{\cal A}}{\lambda_{k+1}}$, and $k:=k+1$, go to Step 1.
\end{description}
\end{Algorithm}

\begin{Remark}\label{rmk-1-a}
By Proposition \ref{prop-2}, if ${\cal A}$ is irreducible and $\lambda\in\Lambda({\cal A})$, then $\lambda>0$. This, together with Proposition \ref{prop-3-a}, implies that Algorithm \ref{algo} is well defined.
\end{Remark}

To establish the convergent of Algorithm \ref{algo}, we prove the following lemma first.

\begin{Lemma}\label{lem-1-a}
Let ${\cal A}\in\mathbb{R}^{m,n}$ be nonnegative, irreducible and symmetric. For $\varrho({\cal A})$, if $\mathbf{x}^*$ is a corresponding Z-eigenvector, then there exists $\epsilon>0$ such that for any $\mathbf{x}^{(0)}\in\mathbb{R}^n_{++}\cap\{\mathbf{x}\in{\cal S}^{n-1}\;|\;\|\mathbf{x}-\mathbf{x}^*\|\leq \epsilon\}$, the sequence $\{\lambda_k\}$ generated by Algorithm \ref{algo1} with $\alpha>(m-1)\varrho({\cal A})$ converges to $\varrho({\cal A})$.
\end{Lemma}

\noindent {\bf Proof.} If $\Lambda({\cal A})$ is the singleton $\{\varrho({\cal A})\}$, then the result follows from Proposition \ref{prop-3-a} with arbitrary $\epsilon>0$. Now, suppose that the cardinality of $\Lambda({\cal A})$ is larger than one.

Denote by $\lambda_2({\cal A}):=\max\{\lambda\;|\;\lambda\in\Lambda({\cal A})\setminus\{\varrho({\cal A})\}\}$, and $\kappa:=\frac{\varrho({\cal A})-\lambda_2({\cal A})}{2}$. By Proposition \ref{prop-2}, we see that the open set $\{\beta\in\mathbb{R}\;|\;|\beta-\varrho({\cal A})|<\kappa\}$ disjoints with the union of the finitely many open sets $\{\beta\in\mathbb{R}\;|\;|\beta-\lambda|<\kappa\}$ for $\lambda\in\Lambda({\cal A})\setminus\{\varrho({\cal A})\}$. Since $\lambda_0:={\cal A}(\mathbf{x}^{(0)})^m$ and $\varrho({\cal A}):={\cal A}(\mathbf{x}^*)^m$, we can choose $\epsilon>0$ such that $|\lambda_0-\varrho({\cal A})|<\kappa$ for any $\mathbf{x}^{(0)}\in\mathbb{R}^n_{++}\cap\{\mathbf{x}\in{\cal S}^{n-1}\;|\;\|\mathbf{x}-\mathbf{x}^*\|\leq \epsilon\}$. Consequently, this, together with Proposition \ref{prop-2}(c), implies that $\varrho({\cal A})-\kappa<\lambda_0\leq\varrho({\cal A})$.

By Proposition \ref{prop-3-a}, $\{\lambda_k\}$ is nondecreasing and converges to a Z-eigenvalue $\lambda_*$ of ${\cal A}$. As we can see, the only possibility is that $\lambda_*=\varrho({\cal A})$. The proof is complete. \ep

\begin{Assumption}\label{assm-a}
If the $N$ initial points in Algorithm \ref{algo} are chosen such that there is at least one point in the set $\{\mathbf{x}\in\mathbb{R}^n_{++}\cap{\cal S}^{n-1}\;|\;\|\mathbf{x}-\mathbf{x}^*\|\leq \epsilon\}$ with $\epsilon$ being determined by Lemma \ref{lem-1-a}, then we call such a set of initial points satisfies {\bf absolutely convergent condition} (ACC for short).
\end{Assumption}

\begin{Theorem}\label{thm-1-a}
Let ${\cal A}\in\mathbb{R}^{m,n}$ be nonnegative, irreducible and symmetric. For $\alpha>(m-1)\varrho({\cal A})$, if the set of $N$ initial points satisfies ACC, then Algorithm \ref{algo} is terminated with $k=1$ and $\lambda=\varrho({\cal A})$. In general, the sequence $\{\lambda_k\}$ generated by Algorithm \ref{algo} converges to a positive Z-eigenvalue of ${\cal A}$.
\end{Theorem}

\noindent {\bf Proof.} The results follow from Proposition \ref{prop-3-a}, Lemma \ref{lem-1-a} and Assumption \ref{assm-a}. \ep

So, if we uniformly randomly choose initial points in $\mathbb{R}^n_{++}\cap{\cal S}^{n-1}$, then with positive probability Algorithm \ref{algo} finds the Z-spectral radius. By Theorem \ref{thm-1-a} and Assumption \ref{assm-a}, we see that the probability that Algorithm \ref{algo} converges to the Z-spectral radius under the uniformly random framework is determined by $\epsilon$ in Lemma \ref{lem-1-a}. In the following subsection, we show that $\epsilon$ in Lemma \ref{lem-1-a} can be determined through $\kappa$ in Lemma \ref{lem-1-a} explicitly.

\subsection{On the ACC assumption}
We in this subsection investigate the $\epsilon$ in Lemma \ref{lem-1-a}. It plays an important role in Assumption \ref{assm-a}, and thus in Theorem \ref{thm-1-a}. We show that there is an explicit formula for $\epsilon$ in Lemma \ref{lem-1-a} based on $\kappa:=\frac{\varrho({\cal A})-\lambda_2({\cal A})}{2}$. For any ${\cal A}=(a_{i_1\ldots i_m})\in\mathbb{R}^{m,n}$, its Frobenius norm $\|{\cal A}\|_F$ is defined as $\|{\cal A}\|_F:=\sqrt{\sum_{i_1,\ldots,i_m=1}^na_{i_1\ldots i_m}^2}$.

\begin{Proposition}\label{prop-4-a}
Let ${\cal A}\in\mathbb{R}^{m,n}$. For any $\mathbf{x},\mathbf{y}\in{\cal S}^{n-1}$, we have
\begin{eqnarray}\label{nin-a}
\|{\cal A}\mathbf{x}^{m-k}-{\cal A}\mathbf{y}^{m-k}\|_F\leq (m-k)\|{\cal A}\|_F\|\mathbf{x}-\mathbf{y}\|,\;\forall k=0,1,\ldots,m.
\end{eqnarray}
\end{Proposition}

\noindent {\bf Proof.} We first show that
\begin{eqnarray}\label{nin-1-a}
\|{\cal A}\mathbf{x}^{(1)}\cdots\mathbf{x}^{(m-k)}\|_F\leq \|{\cal A}\|_F,\;\forall k=0,1,\ldots,m-1.
\end{eqnarray}
Here, $\mathbf{x}^{(1)},\ldots,\mathbf{x}^{(m-k)}\in{\cal S}^{n-1}$.

\begin{itemize}
\item We show how it works for $k=m-1$ in details. In this case, ${\cal A}\mathbf{x}^{(1)}\in\mathbb{R}^{m-1,n}$.
       \begin{eqnarray*}
       \|{\cal A}\mathbf{x}^{(1)}\|^2_F&=&\sum_{i_1,\ldots,i_{m-1}=1}^n\left(\sum_{j=1}^na_{i_1,\ldots,i_{m-1},j}\mathbf{x}^{(1)}_j\right)^2\\
       &\leq& \sum_{i_1,\ldots,i_{m-1}=1}^n\sum_{j=1}^na_{i_1,\ldots,i_{m-1},j}^2\\
       &=&\|{\cal A}\|_F^2.
       \end{eqnarray*}
       Here the inequality follows from the fact that $\mathbf{x}^{(1)}\in{\cal S}^{n-1}$.
\item In general,
\begin{eqnarray*}
       \|{\cal A}\mathbf{x}^{(1)}\cdots\mathbf{x}^{(m-k)}\|_F&=&\|{\cal A}\mathbf{x}^{(1)}\cdots\mathbf{x}^{(m-k-1)}\mathbf{x}^{(m-k)}\|_F\\
       &\leq&\|{\cal A}\mathbf{x}^{(1)}\cdots\mathbf{x}^{(m-k-1)}\|_F\leq\cdots\leq\|{\cal A}\mathbf{x}^{(1)}\|_F\\
       &\leq& \|{\cal A}\|_F
       \end{eqnarray*}
for $k=0,\ldots,m-2$.
\end{itemize}

Now, we show the proof for \reff{nin-a} when $m=3$.
\begin{eqnarray*}
\|{\cal A}\mathbf{x}-{\cal A}\mathbf{y}\|_F&=&\|{\cal A}(\mathbf{x}-\mathbf{y})\|_F=\|P(\mathbf{x}-\mathbf{y})\|\\
&\leq& \|P\|\|(\mathbf{x}-\mathbf{y})\|\leq\|P\|_F\|(\mathbf{x}-\mathbf{y})\|\\
&=&\|{\cal A}\|_F\|(\mathbf{x}-\mathbf{y})\|,
\end{eqnarray*}
here $P$ is an $n^2\times n$ matrix such that $P\mathbf{x}=\mbox{vec}({\cal A}\mathbf{x})$ for all $\mathbf{x}\in\mathbb{R}^n$.
Then, for $m>3$, we have
\begin{eqnarray*}
&&\|{\cal A}\mathbf{x}^{m-k}-{\cal A}\mathbf{y}^{m-k}\|_F\\
&=&\|{\cal A}\mathbf{x}^{m-k}-{\cal A}\mathbf{y}\mathbf{x}^{m-k-1}+{\cal A}\mathbf{y}\mathbf{x}^{m-k-1}-{\cal A}\mathbf{y}^2\mathbf{x}^{m-k-2}+\cdots+{\cal A}\mathbf{y}^{m-k-2}\mathbf{x}-{\cal A}\mathbf{y}^{m-k}\|\\
&\leq&\left(\|{\cal A}\mathbf{x}^{m-k-1}\|_F+\|{\cal A}\mathbf{x}^{m-k-2}\mathbf{y}\|_F+\cdots+\|{\cal A}\mathbf{y}^{m-k-1}\|_F\right)\|(\mathbf{x}-\mathbf{y})\|\\
&\leq&(m-k)\|{\cal A}\|_F\|(\mathbf{x}-\mathbf{y})\|, \;\forall k=0,1,\ldots,m-1.
\end{eqnarray*}
Here, the last inequality follows from \reff{nin-1-a}.  The proof is complete. \ep
\begin{Corollary}\label{cor-4-a}
Let ${\cal A}\in\mathbb{R}^{m,n}$ be nonnegative, irreducible and symmetric. For any $\mathbf{x}^{(0)}\in\mathbb{R}^n_{++}\cap\{\mathbf{x}\in{\cal S}^{n-1}\;|\;\|\mathbf{x}-\mathbf{x}^*\|< \frac{\kappa}{m\|{\cal A}\|_F}\}$, we have $|\lambda_0-\varrho({\cal A})|<\kappa$.
\end{Corollary}
\begin{Remark}\label{rmk-2-a}
By Theorem \ref{thm-1-a} and Corollary \ref{cor-4-a}, we see that $\kappa$ determines the probability that Algorithm \ref{algo} converges to $\varrho({\cal A})$ under the random framework. Note that $\|{\cal A}_{\Psi}\|_F=1$ for pure states. Then, we can analyze the probability through estimations of $\kappa$ for computing the geometric measure of entanglement.
\end{Remark}

\end{document}